\documentclass[aps,prb,twocolumn,showpacs,amsmath,amssymb]{revtex4}

\usepackage{graphicx}
\usepackage{epsfig}

\begin{document}

\title{The short-time behavior of a classical ferromagnet with
double-exchange interaction} \author{H. A. Fernandes and
J. R. Drugowich de Fel\'{\i}cio} \affiliation{Departamento de
F\'{\i}sica e Matem\'{a}tica, Faculdade de Filosofia, Ci\^{e}ncias e
Letras de Ribeir\~{a}o Preto, Universidade de S\~{a}o Paulo, Avenida
Bandeirantes, 3900 - CEP 14040-901, Ribeir\~{a}o Preto, S\~{a}o Paulo,
Brazil} \author{A. A. Caparica\footnote{Corresponding author:\\
caparica@if.ufg.br (A. A. Caparica)}} \affiliation{Instituto de
F\'{\i}sica, Universidade Federal de Goi\'{a}s, 74.001-970,
Goi\^{a}nia (GO) Brazil}

\begin{abstract}
We investigate the critical dynamics of a classical ferromagnet on the
simple cubic lattice with double-exchange interaction. Estimates for
the dynamic critical exponents $z$ and $\theta$ are obtained using
short-time Monte Carlo simulations. We also estimate the static
critical exponents $\nu$ and $\beta$ studying the behavior of the
samples at an early time. Our results are in good agreement with
available estimates and support the assertion that this model and the
classical Heisenberg model belong to the same universality class.
\end{abstract}


\maketitle

\section{INTRODUCTION}
In the last decade many works have been devoted to the study of the
perovskite manganites, $A_{1-x}B_xMnO_3$, where $A$ and $B$ are
rare-earth and alkaline-earth ions, respectively, and $x$ is the
concentration of $B$. The perovskite manganites are metal oxides that
exhibit some remarkable properties. The most noticeable of them is
colossal magnetoresistance (CMR), an extremely large change in the
resistivity when a magnetic field is applied in the vicinity of the
critical temperature $T_c$.  Usually these materials have been studied
on the basis of the double-exchange theory
\cite{Zener1951,Anderson1955,Gennes1960,Varma1996, Furukawa1994}
(DE). In one of these works, Anderson e Hasegawa \cite{Anderson1955}
showed that the transference element is proportional to
$cos(\theta/2)$, where $\theta$ is the angle between the ionic
neighbouring spins. This result was recently confirmed for layered
manganites \cite{Qingan2000}. Nevertheless, although this theory has
succeeded in explaining qualitatively CMR, some authors have argued
that it cannot alone provide a complete description of this
phenomenon.  They suggest that, in addition to the double exchange, a
complete understanding of these materials should include strong
electron correlations \cite{Asamitsu1996}, a strong electron-phonon
interaction \cite{Millis1995}, or coexisting phases
\cite{Lynn1996}. One might therefore think that double-exchange alone
cannot explain CMR in manganites \cite{Millis1995}, but this remains
an open question. What we know is that in the study of the manganites
the double-exchange theory plays an important role, both in the study
of CMR and in explaining the presence of a ferromagnetic state (for $x
\thickapprox 0.3$) in doped manganites, furnishing the basis for
describing manganites with colossal magnetoresistence.

The mechanism of the double-exchange consists of a strong intrasite
exchange between localized ions (core spin $S_c^i$) and non-localized
electrons ($E_g$ orbitals). Based on this hypothesis Zener
\cite{Zener1951} succeded in explaining in 1951 the occurrence of the
metal ferromagnetic phase in manganites. Due to a strong intrasite
exchange interaction the spin of the electron always aligns parallel
to the spin of the ion.  When an electron hops from site $i$ to site
$j$ its spin must also change from being parallel to $\textbf{S}_c^i$
to being parallel to $\textbf{S}_c^j$, but with an energy loss
proportional to the cosine of half the angle between the core spins
\cite{Anderson1955}.

In Ref. [10], the critical behavior of a classical ferromagnet with
double-exchange interaction was studied and estimates for the static
critical exponents $\nu$, $\beta$ and $\gamma$, and the critical
temperature $T_c$ were derived. These simulations indicate that this
model belongs to the universality class of the classical Heisenberg
model \cite{Chen1993}.

In this paper, we perform short-time Monte Carlo simulations to
explore the critical dynamics of a classical ferromagnet with
double-exchange interaction. We evaluate the dynamic exponents $z$,
$\theta$ and the static exponents $\nu$ and $\beta$. Our estimates
for static indices are in good agreement with previous results
available in the literature \cite{Caparica2000,Motome2000,Motome2001}
for the same model. In addition, the result for the dynamic exponent
$z$ compare well with previous estimates for the classical Heisenberg
model. In summary, present results confirm that both models belong to
the same universality class.

The article is organized as follows. In the next Section we define the
model. In Section III we give a brief description of short-time Monte Carlo
technique and present our results; in Section IV we summarize and
conclude.

\section{THE MODEL}

We use the short-time Monte Carlo simulations to investigate the
dynamical critical behavior of a classical spin model with double-exchange
interaction. We consider a simple cubic lattice $L \times L \times L$
with periodic boundary conditions. The Hamiltonian for such system is
given by \cite{Caparica2000}
\begin{equation}
\mathcal{H}=-\sum_{<i,j>}J\sqrt{1+\textbf{S}_i\cdot\textbf{S}_j} \label{hde},
\end{equation}
where $<i,j>$ indicates that the sum runs over all nearest-neighbor
pairs of lattice site, $J$ is the ferromagnetic coupling constant and
the spin $\textbf{S}_i=(S_i^x,S_j^y,S_k^z)$ is a three-dimensional
vector of unit length.

We used lattice sizes $L=20$, $25$, $30$, $40$ and $60$ and performed
short-time simulations at the critical temperature \cite{Caparica2000}
\begin{equation*}
T_c=0.74515,
\end{equation*}
in units of $J/k_B$, where $k_B$ is the Boltzman's constant.

Our estimates for each exponent were obtained from five independent
bins. The dynamic evolution of the spins is local and updated by the
Metropolis algorithm. For $20\geq L \geq40$ the estimate for the exponent
$\theta $ was performed using 10000 independent samples, while for
the exponents $z$, $\nu $ and $\beta $ we used 5000 samples. For
$L=60$ all the estimates were performed using 2000 samples. In order
to maximize the quality of the linear fitting, each exponent is
determined in the time interval $[200,500]$, except for the exponent
$z$ which was measured in the interval $[30,250]$.

\section{SHORT-TIME SCALING RELATIONS AND RESULTS}

Following the works of Janssen, Schaub and Schmittmann \cite{Janssen1989}
and Huse \cite{Huse1989}, the study of the critical
properties of statistical systems became in some sense simpler,
because they allow to circumvent the well-known problem of the
critical slowing down, characteristic of the long-time regime.
They discovered using renormalization group techniques and numerical
calculations, respectively, that there is universality and scaling
behavior even at the early stage of the time evolution of dynamical
systems without conserved quantities (model A in the terminology of
Halperin \textit{et al} \cite{Halperin1974}).

In this new universal regime, in addition to the familiar set of
static critical exponents and the dynamic critical exponent $z$, a new
dynamic critical exponent $\theta$ is found.  This new critical index,
independent of the previously known exponents,characterizes the
so-called \textit{critical initial slip}, the anomalous increase of
the magnetization when the system is quenched to the critical
temperature $T_c$.

Several works on phase transitions and critical phenomena
using this technique have been published \cite{Zheng1998}. The results 
corroborate the prediction of universality and scaling behavior 
in the short-time regime and the static critical exponents so obtained
are in good agreement with well-known values calculated in the 
equilibrium. In addition, this approach has permitted obtaining
more precise estimates for the dynamic critical exponent $z$.

The dynamic scaling relation obtained by Janssen \textit{et al} for
the \textit{k}-th moment of the magnetization, extended to systems
of finite size \cite{Li1995}, is written as
\begin{equation}
M^{(k)}(t,\tau ,L,m_{0}) =b^{-k\beta /\nu}M^{(k)}(b^{-z}t,b^{1/\nu}\tau
,b^{-1}L,b^{x_{0}}m_{0}) \label{eq1},
\end{equation}
where $t$ is time, $b$ is an arbitrary spatial rescaling
factor, $\tau=\left(T-T_{c}\right)/T_{c}$ is the reduced temperature
and $L$ is the linear size of the lattice. The exponents $\beta$
and $\nu$ are the equilibrium critical exponents associated with the
order parameter and the correlation length, and $z$ is the dynamic
exponent characterizing time correlations in equilibrium. For a
large lattice size $L$ and small initial magnetization $m_0$ at the
critical temperature $(\tau=0)$, the magnetization is governed by a
new dynamic exponent $\theta$,
\begin{equation}
M(t) \thicksim m_0t^\theta \label{eq2}
\end{equation}
if we choose the scaling factor $b=t^{1/z}$.

In addition, a new critical exponent $x_{0}$, which represents the
anomalous dimension of the initial magnetization $m_{0}$, is
introduced to describe the dependence of the scaling behavior on the
initial conditions. This exponent is related to $\theta$ as
$x_0=\theta z+\beta/\nu$.  In the following section
 we evaluate the critical exponents $\theta$, $z$,
$\nu$ e $\beta$ using scaling relations obtained from Eq. (\ref{eq1}).

\subsection{The dynamic exponent $\theta$}

Usually the dynamic critical exponent $\theta$ is calculated using
Eq. (\ref{eq2}) or through the autocorrelation
\begin{equation}
A(t) \thicksim t^{\theta-\frac{d}{z}}, \label{auto}
\end{equation}
where $d$ is the dimension of the system.

In the first case, careful preparation of the initial configurations
($m_0 \ll 1$) is essential, as well as the delicate limit $m_0
\rightarrow 0$. In the autocorrelation case, one must to know in
advance the exponent $z$, which is an order of magnitude greater
than $\theta$, so that a small relative error in $z$ induces a large
error in $\theta$.

In the present work we find the dynamic critical exponent $\theta$
using the time correlation of the magnetization \cite{Tania1998}
\begin{equation}
C(t)=\langle M(0)M(t)\rangle \thicksim t^{\theta}. \label{teta}
\end{equation}
Here, the averaging is over a set of random initial
configurations. The time correlation allows the direct calculation of
the dynamic exponent $\theta$, without the need of a careful
preparation of the initial state nor of the limiting procedure, the
only requirement being that $\langle M_0 \rangle=0$.

\begin{figure}[ht]
\centering \epsfig{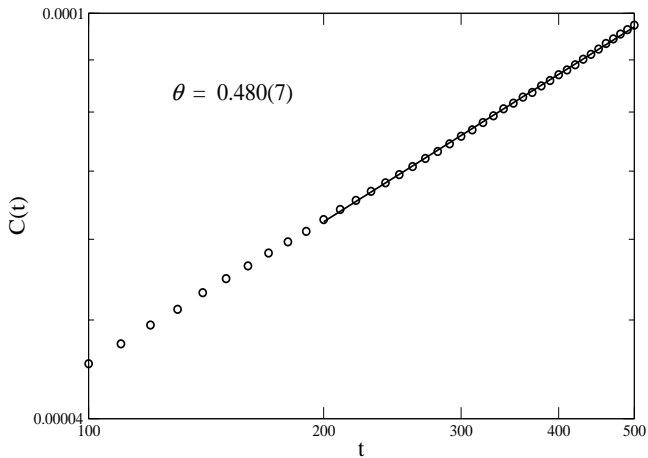}
\caption{Time evolution of the time correlation of the total
magnetization on log scales. Errorbars are smaller than the symbols.
Each point represents an average over 5 sets of 2000 samples.}
\label{fig:theta}
\end{figure}

In Fig. \ref{fig:theta} we show the time dependence of the time
correlation $C(t)$ in double-log scale for the system with $L=60$.
Table \ref{table:exponents} shows our estimates for the critical exponent
$\theta$ along with $d/z$, $1/\nu z$ and $\beta/\nu z$.
\begin{table}[!ht]\centering
\caption{Critical exponents for the double-exchange model.}\label{table:exponents}
\begin{tabular}{c c c c c}
\hline\hline $L$ & $\theta$ & $d/z$ & $1/\nu z$ & $\beta/\nu z$ \\
\hline 20  & 0.460(4) & 1.499(1) & 0.766(6) & 0.276(3) \\
       25  & 0.473(4) & 1.515(2) & 0.734(5) & 0.267(3) \\
       30  & 0.479(8) & 1.521(5) & 0.748(7) & 0.267(2) \\
       40  & 0.482(4) & 1.525(7) & 0.743(5) & 0.264(1) \\
       60  & 0.480(7) & 1.519(8) & 0.738(7) & 0.263(2) \\
\hline\hline
\end{tabular}
\end{table}

These results show that for $L \geq 25$ the finite size effects are
less than the statistical errors. Therefore, we conclude that the values
for an infinite lattice are within the errorbars of our results for $L=60$.

\subsection{The dynamic exponent $z$}

In the short-time regime the dynamic critical exponent $z$ can be
evaluated using various methods, such as the autocorrelation \cite{Schulke1995}
(Eq. (\ref{auto}), with the previous knowledge of $\theta$), or the second
moment $M^{(2)}(t)$ \cite{Zheng1998} (when the static exponents
$\beta$ and $\nu$ are known). In the present work we estimate the
exponent $z$ in two independent ways, using two different techniques.
First, we used the time-dependent fourth-order Binder cumulant \cite{Binder1981,Li1995}
\begin{equation}
U_4(t,L)=1-\frac{\langle M^4 \rangle}{3 \langle M^2 \rangle ^2}, \label{cum}
\end{equation}
with ordered initial configurations $(m_0=1)$. If $\tau=0$ this equation
depends only on $t/L^z$, according to scaling laws valid in the
beginning of the evolution.

In this case, according to the scaling relation Eq. (\ref{eq1}),
the cumulants obtained in two different lattices obey the equation
\begin{equation}
U_4(t,L_1)=U_4(b^{-z}t,L_2)
\end{equation}
with $b=L_1/L_2$. This equation shows that the exponent $z$ is
obtained by finding the time re-scaling factor $b^{-z}$ which colapses
curves for two lattices of different sizes.

\begin{figure}[ht]
\centering \epsfig{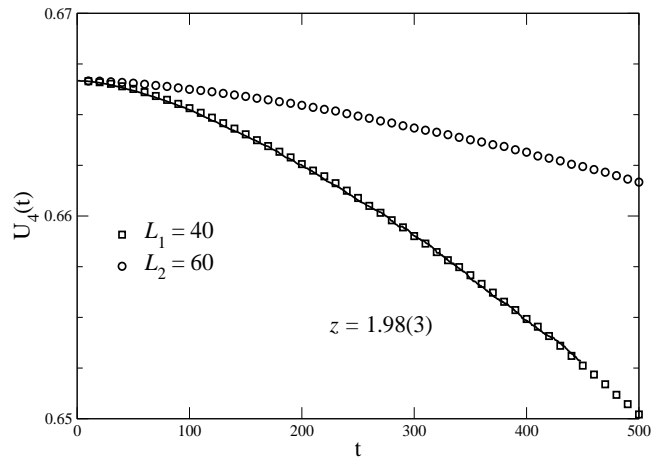} \caption{The
cumulant $U_4(t,L)$ for $L=40$ and $60$ plotted versus time $t$ with
initial states $m_0=1$. The line shows the lattice $L=60$ re-scaled
in time. The cumulant for both lattice sizes was obtained from five
independent bins. For $L=40$ we used 5000 samples and for $L=60$ we
used 2000 samples.} \label{fig:cum}
\end{figure}

In Fig. \ref{fig:cum} we show the Binder cumulant as a function of
the time for the lattices $L_1=60$ and $L_2=40$.
The line shows the colapse of the larger lattice re-scaled in time.
Our estimate for $z$, obtained through the $\chi^2$ test
\cite{Numerical}, for $L_1=60$ and $L_2=40$ is
\begin{equation}
z=1.98(3) \label{eq:cum}
\end{equation}

This result is consistent with that obtained by Peczak and Landau
\cite{Landau1990} for the classical Heisenberg model. Using
equilibrium tecniques, these authors found $z=1.96(7)$. We also
estimate the exponent $z$ for other lattice sizes and our results
are consistent with the Eq. (\ref{eq:cum}). For $L_1=40$ and
$L_2=25$ we obtained $z=1.96(2)$, whereas for $L_1=60$ and $L_2=30$
we obtained $z=1.96(4)$.

A more precise estimate for the dynamic exponent $z$ can be obtained
by combining results from samples submitted to different initial
conditions ($m_0=1$ and $m_0=0$). According to the Ref. [24] the
function $F_2(t)=M^{(2)}(t)/(M(t))^2$ behaves as $t^{d/z}$ where $d$ is the
dimension of the system. This approach proved to be very efficient in
estimating the exponent $z$, according to results for the Ising model,
the three- and four-state Potts models \cite{Silva20021}, the
tricritical point of the Blume-Capel model\cite{Silva20022}, the
Baxter-Wu model \cite{Arashiro2003} and nonequilibrium models like
Domany-Kinzel and contact process \cite{Silva2004}. 
In this technique, for different lattice
sizes, the double-log curves of $F_2$ \textit{versus} $t$ fall on a
single straight line, without any rescaling of time, resulting in more
precise estimates for $z$.

\begin{figure}[ht]
\centering
\epsfig{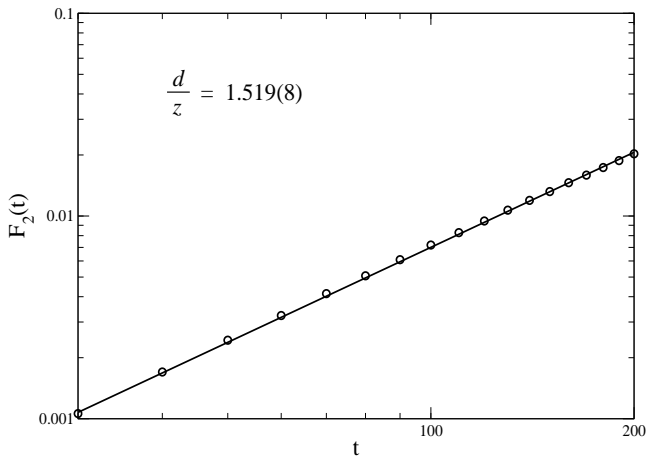}
\caption{Time Evolution of $F_2$. The error bars, calculated over 5
sets of 2000 samples, are smaller than the symbols.} \label{fig:f2}
\end{figure}

The time evolution of $F_2$ is shown on log scales in
Fig. \ref{fig:f2} for $L=60$.
In Table \ref{table:exponents} we show the values of $d/z$ for the
five lattice sizes. Taking into account the value of this ratio for
the lattice $L=60$ we obtain
\begin{equation}
z=1.975(4) \label{eq:f2}
\end{equation}
Our results for $z$ indicate that besides presenting the same static
critical exponents double-exchange and Heisenberg models have the same
dynamic exponent when submitted to the same updating.

\subsection{The static exponents $\nu$ and $\beta$}

\begin{figure}[ht]
\centering
\epsfig{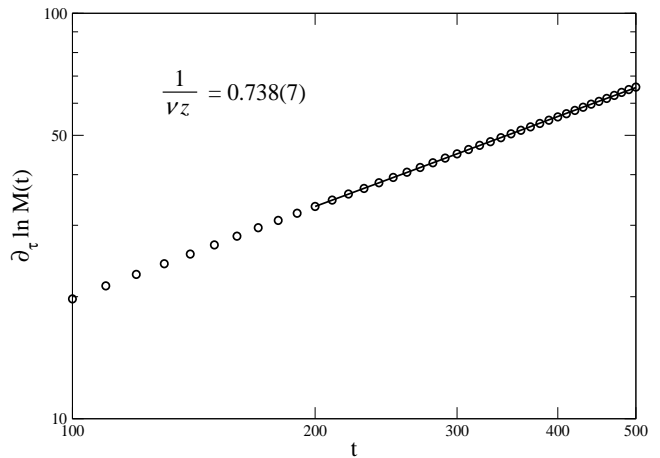}
\caption{The time evolution of the derivative $\partial_\tau
\mbox{ln}M(t,\tau)|_{\tau=0}$ on log scales in a dynamic process
starting from an ordered state $(m_0=1)$. Error bars are smaller
than the symbols. Each point represents an average over 5 sets of 5000
samples.} \label{fig:dvd}
\end{figure}

The static exponent $\nu$ can be obtained by fixing $b^{-z}t=1$
in Eq. (\ref{eq1}) and differentiating $\mbox{ln} M(t,\tau)$ with respect to
$\tau$ in the critical point. Taking into account samples with ordered
initial configurations $(m_0=1)$, we obtain the following power law
\begin{equation}
\partial_\tau \mbox{ln}M(t,\tau)|_{\tau=0} \thicksim t^{1/\nu z} \label{dm}.
\end{equation}
In numerical simulations we approximate the derivative by a finite
difference; our results were obtained using finite differences of
$T_c\pm \delta$ with $\delta=0.01$, in units of $J/k_B$. In
Fig. \ref{fig:dvd} the power law increase of Eq. (\ref{dm}) is plotted
in double-log scale for $L=60$.

From the slope of the curve one can estimate the critical exponent
$1/\nu z$ (see Table \ref{table:exponents}).

Using the exponent $z$ obtained from $U_4(t,L)$ and the estimate of
$1/\nu z$ for $L=60$, we obtain the static exponent
\begin{equation}
\nu=0.68(2).
\end{equation}
while following the scaling relation $F_2(t)$ the result is
\begin{equation}
\nu=0.682(8).
\end{equation}

\begin{figure}[ht]
\centering
\epsfig{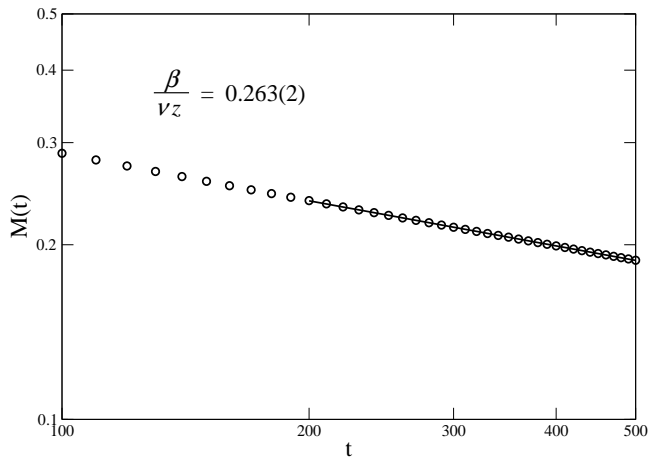}
\caption{The time evolution of the magnetization for initially ordered
samples $(m_0=1)$. The errorbars calculated over 5 sets of 2000
samples, are smaller than the symbols.} \label{fig:mag}
\end{figure}

Finally we can evaluate the static exponent $\beta$ by the
dynamic scaling law for the magnetization
\begin{equation}
M(t) \thicksim t^{-\beta/{\nu z}} \label{mag},
\end{equation}
obtained from Eq. (\ref{eq1}), by considering large systems and setting
$b=t^{1/z}$ at the critical temperature $\tau=0$. In Fig. \ref{fig:mag} we show
the time evolution of the magnetization in double-log scale for $L=60$.
The estimates of the exponent $\beta/\nu z$ for the five lattice
sizes are shown in Table \ref{table:exponents}.

Using the previous result obtained for $1/\nu z$ (Table \ref{table:exponents}), we find
\begin{equation}
\beta=0.354(5).
\end{equation}

Our estimates for the static exponents $\nu$ and $\beta$ are in good
agreement with the theoretical results of the Ref. [10]
($\nu=0.6949(38)$ and $\beta=0.3535(30)$), Ref. [11] ($\nu=0.7036(23)$
and $\beta=0.3616(31)$) and Ref. [28] ($\nu=0.704(6)$ and
$\beta=0.362(4)$), as well as the experimental results of
the Ref. [29] ($\beta=0.37(4)$) and the Ref. [30] ($\beta=0.374(6)$).

\section{DISCUSSION AND CONCLUSIONS}

In this article we have performed short-time Monte Carlo simulations
in order to evaluate the dynamic and static critical exponents of a
classical ferromagnet with double-exchange interaction. The exponent
$\theta$ was estimated using the time correlation of the
magnetization. We found the dynamic exponent $z$ through the colapse
of the fourth-order time-dependent Binder cumulant and also,
alternatively, using the function $F_2(t)$ which combines
simulations performed with different initial conditions. Using
scaling relations for the magnetization and its derivative with
respect to the temperature at $T_c$, we have obtained the static
exponents $\nu$ and $\beta$ which are in good agreement with the values
available in literature. Comparison of our results with those for
the classical Heisenberg model corroborate the assertion that the
models belong to the same universality class.

\section*{ACKNOWLEDGMENTS}

The authors are grateful to Departamento de F\'{i}sica e Matem\'{a}tica (DFM)
at IFUSP for the use of their computer facilities. This work was supported
by the Brazilian agencies CAPES (H.A.F) and FUNAPE-UFG (A.A.C.).

\end{document}